\documentclass[aps,a4paper,pre,showpacs,floatfix,superscriptaddress,twocolumn,merge,longbibliography,elide]{revtex4-1}
\usepackage{graphicx}
\usepackage[dvipsnames]{pstricks}
\usepackage{amsmath,amsfonts,amssymb}
\def\bg#1{\hbox{\bf#1}}
\def\sg#1{\hbox{\sf#1}}
\newcommand{\be}{\begin{eqnarray}}
\newcommand{\ee}{\end{eqnarray}}
\newcommand{\bC}{\begin{center}}
\newcommand{\eC}{\end{center}}
\newcommand{\befi}{\begin{figure}}
\newcommand{\enfi}{\end{figure}}

\newcommand{\ci}{\cite}
\newcommand{\la}{\label}

\newcommand{\benl}{\begin{eqnarray*}}
\newcommand{\eenl}{\end{eqnarray*}}
\newcommand{\REM}[1]{}

\usepackage[%
  unicode=true,%
  colorlinks=true,%
  linkcolor=blue,anchorcolor=blue,%
  breaklinks=true,%
  citecolor=blue,filecolor=cyan,%
  menucolor=blue,%
  urlcolor=blue]{hyperref}

\begin{document}
\title{Nematic order in a simple--cubic lattice--spin model
\\
with full--ranged dipolar interactions
}

\author{Hassan Chamati}
\affiliation{Institute of Solid State Physics, Bulgarian Academy of Sciences,
72 Tzarigradsko Chauss\'ee, 1784 Sofia, Bulgaria\\
chamati@bas.bg  }
\author{Silvano Romano}
\affiliation{Physics dept., the University,
via A. Bassi 6, 27100 Pavia, Italy \\
silvano.romano@pv.infn.it}

\begin{abstract}
In a previous paper
[\href{http://dx.doi.org/10.1103/PhysRevE.90.022506}{
Phys. Rev. E 90, 022506 (2014)}], we had studied
thermodynamic and structural properties of a three--dimensional
simple--cubic
lattice model with dipolar--like interaction, truncated at
nearest--neighbor separation, for which the existence of an ordering
transition at finite temperature had been proven mathematically;
here we extend our investigation addressing the
full--ranged counterpart of the model, for which the critical
behavior had been investigated theoretically and experimentally. In
addition the existence of an
ordering transition at finite temperature had been proven mathematically
as well. Both models
exhibited the same continuously degenerate ground--state
configuration, possessing full orientational order with respect to a
suitably defined staggered magnetization (polarization), but no
nematic second--rank order; in both cases, thermal fluctuations remove
the degeneracy, so that nematic order does set in at low but finite
temperature via a mechanism of order by disorder. On the other hand,
there were recognizable quantitative differences between the two
models as for ground--state energy and critical exponent estimates;
the latter were found to agree with early Renormalization Group 
calculations and with experimental results.
\end{abstract}
\date{\today}
\maketitle

\section{Introduction} \label{intro}
Long--range dipolar interactions \footnote{Both electrostatic and
magnetostatic dipolar interactions have the same mathematical
structure (within numerical factors and usage of different units or
symbols), and both interpretations are currently used in the
Literature; in the following, we shall be using the magnetic
language.} between magnetic moments are ubiquitous in experimentally
studied magnetic systems, although often dominated by exchange
couplings (for more details see Refs.
\cite{debell2000,kaul2005,shand_magnetic_2008,ravi_kumar_magnetic_2015} and
references therein), and, over decades, a number of theoretical
studies, based on Renormalization group techniques, 
has addressed interaction models containing both dipolar and short--range
isotropic or anisotropic exchange interactions (see, {\it e.g.}, Refs
\cite{bruce_critical_1974,frey_renormalized_1991,ried_phase_1995,debell2000,meloche_dipole-exchange_2011,mol2014,wysin_order_2015}); on the other hand, 
lattice models involving only the long--range dipolar term have also
 long been studied by various approaches,
 including spin-wave treatments and simulation
 (see, \textit{e.g.}, Refs.
\cite{jetp05700636,PhysRev.99.1135,mp02801359,mp03200931,pre90022506,prb91020410,holden_monte_2015,johnston_magnetic_2016},  and others quoted in the following). While the former references have dealt with the 
resulting critical behavior, 
including the crossover between isotropic dipolar
universality class (when the dipolar term is dominant) and 
the Heisenberg one (corresponding to nearest-neighbor exchange interactions
only),
the later ones were \textit{mostly} focused
on the ground state of the magnetically ordered phase; in addition, a survey of
relevant rigorous mathematical results can be found in \cite{jsp1341059}. In a
previous paper \cite{pre90022506}, we had studied thermodynamic and
structural properties of a three--dimensional lattice model with
dipolar--like interaction, truncated at nearest--neighbor separation,
for which the existence of an ordering transition at finite
temperature had been proven mathematically \cite{rFroh}. It was found
that the ground state is degenerate and the critical
behavior of the model is consistent with the Heisenberg universality
class; moreover, the model was found to exhibit a nematic order
induced by thermal disorder; the study of an \textit{isolated} cubic
dipole cluster \cite{prb91020410} was published shortly
afterwards by other Authors, where the degeneracy of the ground state
was found in this case as well.

Some similar studies have been published only recently in Refs.
\cite{holden_monte_2015,johnston_magnetic_2016}, and have
addressed the ground state of a model of pure dipolar interaction
considering different types of lattices; the magnetic properties of
the ground state were determined for each lattice structure.

Among the above studies, only Ref. \cite{pre90022506} investigating a
pure dipolar model with interaction restricted to nearest-neighbor pairs of
sites have considered the possibility of nematic ordering both in the
ground state and at finite temperature. Here we continue addressing
the full--ranged counterpart of the model, for which mathematical
results have been produced \cite{jsp1341059} as well;
the treatment was based on Reflection Positivity \cite{rFILS},
and proved 
the existence of an ordering transition at finite temperature,
as predicted by spin--wave theory.

The interaction model studied here has long-range tails  expected to
alter the critical behavior of its counterpart  investigated in Ref.
\cite{pre90022506}. Our results
show a downward shift of the critical temperature,
and, in addition, lead 
to different values of critical exponents, as well as critical
amplitudes, thus pointing to a class of universality beyond the
nearest--neighbor Heisenberg one.
We are also revisiting and correcting an earlier and
crude simulation study of the full--ranged model,
carried out by one of us some thirty years ago
\cite{nuovocimd7,epl2}.

The rest of our paper is organized as follows: 
in Section \ref{ground} 
results for the ground state of interaction potential 
\eqref{e01} are recalled;
the simulation methodology is briefly discussed in Section 
\ref{comptaspect}; simulation results 
and  Finite--Size Scaling analysis are used in Section \ref{results} 
to extract the critical 
behavior for the model under consideration. The paper is concluded by a
Section \ref{conclusions}, where  results are summarized.

\section{Interaction Model  and  Ground State} \label{ground}
In keeping with our previous work 
\cite{prb4912287,pre90022506}, we are considering a
classical system consisting of $n-$component magnetic
moments to be denoted by unit vectors $\{ \mathbf{u}_j \}$,
with orthogonal Cartesian components $u_{j,\iota}$,
defined with respect to lattice axes,
associated with a $d-$dimensional lattice 
$\mathbb{Z}^d$ (here $d=n=3$), and interacting 
via a translationally invariant pair potential of the 
form
\begin{equation}
V_{ij} = \epsilon f(r) [-3 (\mathbf{u}_i \cdot \hat{\mathbf{r}}_{ij})
	(\mathbf{u}_j \cdot \hat{\mathbf{r}}_{ij}) + \mathbf{u}_i \cdot 
	\mathbf{u}_j ],~
\label{e01}
\end{equation}
with $\epsilon$  a positive quantity setting energy 
and temperature scales (i.e. energies will be expressed
in units of $\epsilon$, and temperatures defined by
$T=k_B \, \mathcal{T}_K/\epsilon$, where $\mathcal{T}_K$ denotes the 
temperature in degrees Kelvin), and
$$
\mathbf{r}_{ij}=\mathbf{x}_i - \mathbf{x}_j,~
r = |\mathbf{r}_{ij}|,~\hat{\mathbf{r}}_{ij}= \frac{\mathbf{r}_{ij}}{r},
~f(r) > 0;
$$
here $\mathbf{x}_j$ denotes  dimensionless lattice site 
coordinates, and now
$f(r)=r^{-3}$; in Ref. \cite{pre90022506}
the interaction was restricted to nearest neighbor separations
\textit{i.e} $f(r) = 1$ for $r = 1$ and $0$ otherwise.

Both the full--ranged and the nearest--neighbor counterpart possess the same 
continuously degenerate ground--state
configuration (see also below); the ground--state energies (in units $\epsilon$
per particle) are $W_{GS}=-4$ for the nearest--neighbor model
\cite{prb4912287,pre90022506}
and $W_{GS}=-2.676$ for the present full--ranged counterpart
\cite{pre90022506,prb4912287,jetp05700636}.
Extensive references to
Luttinger--Tisza methodologies for
ground--state calculation can be also be found in
\cite{johnston_magnetic_2016}.

For the sake of clarity and completeness we recall here some properties of 
the continuously degenerate ground state
for the three--dimensional case ($d=3$),  closely
following the corresponding section in our previous paper
\cite{pre90022506}, Let lattice site coordinates be expressed as
${\bg x}_j={\bg x}(h,k,l)  =  h {\bg e}_1 + k {\bg e}_2+l{\bg e}_3,~d=3$,
where ${\bg e}_{\alpha}$ denotes unit vectors along the 
lattice axes;
here the subscript in $h_j$ has been omitted for ease
of notation; let also $\varrho_h=(-1)^h$, $\sigma_{hk}=\varrho_h 
\varrho_k$, $\tau_{hkl}=\varrho_h \varrho_k \varrho_l$.;
the ground state possesses continuous degeneracy,
and the manifold of its possible configurations is
defined by \cite{jetp05700636}
\begin{eqnarray} 
{\bg u}^0_j = {\bg u}^0(h,k,l)=
\sigma_{kl} N_1 {\bg e}_1 + \sigma_{hl} N_2 {\bg e}_2 +
\sigma_{hk} N_3 {\bg e}_3,
\label{e05}
\end{eqnarray}
where
\begin{subequations}
\label{mvect}
\begin{align}
N_1 = & \sin \Theta \cos \Phi,
\label{mvx}
\\
N_2 = & \sin \Theta \sin \Phi,
\label{mvy}
\\
N_3 = & \cos \Theta,
\label{mvz}
\end{align}
\end{subequations}
and
$0 \le \Theta \le \pi,~0 \le \Phi \le 2 \pi$;
we also found  it advisable to use the superscript $0$
for various ground--state quantities;
the above configuration will be 
denoted by $D\left( \Theta,~\Phi \right)$.

Various structural quantities can be defined, 
some of which are found to be zero for all values of 
$\Theta$ and $\Phi$, or to average to zero 
upon integration over the angles; for example,  when  $d=3$,
\begin{subequations}
\begin{align}
\sum_{j \in \Delta} {\bg u}^0_j &= 0,~ \\
\sum_{j \in \Delta} \rho_h {\bg u}^0_j &= 0,~ 
\sum_{j \in \Delta} \rho_k {\bg u}^0_j = 0,~
\sum_{j \in \Delta} \rho_l {\bg u}^0_j = 0,~ \\
\sum_{j \in \Delta} \tau_{hkl} {\bg u}^0_j &= 0.
\end{align}
\label{eqsumto0}
\end{subequations}
Here $\Delta$ denotes the $d-$dimensional 
unit cell, and $\varrho= 2^d$ the number of particles in it;
other staggered magnetizations
are not averaged to zero upon summing over the unit cell:   
\begin{subequations}
\label{Cvect}
\begin{align}
{\bg B}_1^0 &=  \sum_{j \in \Delta} 
\sigma_{kl} {\bg u}^0_j = \varrho N_1 {\bg e}_1,~
\label{b1}
\\
{\bg B}_2^0 &=  \sum_{j \in \Delta} 
\sigma_{hl} {\bg u}^0_j = \varrho N_2 {\bg e}_2,~
\label{b2}
\\
{\bg B}_3^0 &=  \sum_{j \in \Delta} 
\sigma_{hk} {\bg u}^0_j = \varrho N_3 {\bg e}_3;~
\label{b3}
\end{align}
\end{subequations}
thus, bearing in mind the above formulae,  for any
unit vector $\mathbf{u}_j$ associated with the lattice
site $\mathbf{x}_j$, one can define another unit vector $\mathbf{w}_j$
with Cartesian components $w_{j,\kappa}$ via
\begin{subequations}
\label{wblock}
\begin{align}
w_{j,1} &  = \sigma_{kl} u_{j,1}
\label{wx}
\\
w_{j,2} & = \sigma_{hl} u_{j,2}
\label{wy}
\\
w_{j,3} & = \sigma_{hk} u_{j,3}
\label{wz}
\end{align}
\end{subequations}
and hence the staggered magnetization
\begin{equation}
\mathbf{C} = \sum_{j \in \Delta} \mathbf{w}_j; 
\label{cvecnew}
\end{equation}
when $\mathbf{u}_j=\mathbf{u}^0_j,~j=1,2 \ldots 8$,
i.e. for the ground--state orientations, Eqs. \eqref{e05} and  
\eqref{cvecnew} lead to
\begin{equation}
{\bg C}^0 = \sum_{j \in \Delta} \mathbf{w}^0_j =  
{\bg B}_1^0 + {\bg B}_2^0 + {\bg B}_3^0 =
\varrho \left(N_1 {\bg e}_1 + N_2 {\bg e}_2 + N_3 {\bg e}_3\right);
\label{e11}
\end{equation}
in this case
\begin{equation}
{\bg w}^0_j = N_1 {\bg e}_1 + N_2 {\bg e}_2 + N_3 {\bg e}_3,~j=1,2 \ldots 8 .
\label{e11-a}
\end{equation}
The ground--state order parameter is defined by
\begin{equation}
\frac1{\varrho}\sqrt{{\bg C}^0 \cdot {\bg C}^0} = 1.
\label{e11-b}
\end{equation}
Eqs. (\ref{mvect}), (\ref{e11}) and (\ref{e11-a}) show that
in all $D\left( \Theta,~\Phi \right)$ configurations
the vector $\mathbf{C}^0$ has the same modulus, and that
each $D(\Theta,\Phi)$ defines its possible
orientation, or, in other words,
the ground state exhibits full order and continuous degeneracy
with respect to the above $\mathbf{C}^0$ vector.
Notice also that the above transformation from   $\mathbf{u}_j$
to $\mathbf{w}_j$ unit vectors (Eq. (\ref{wblock})) can, and will 
be, used in the following for arbitrary configurations of 
unit vectors $\mathbf{u}_j$, 
to calculate $\mathbf{C}$ (Eqs. (\ref{cvecnew})) and related quantities.

As for nematic ordering in the ground state.
for a generic configuration $D\left( \Theta,\Phi \right)$,
the nematic second--rank ordering tensor 
${\sg Q}^0$ is defined by \cite{r23,r24,rbpz-02}
\begin{equation}
Q_{\iota\kappa}^0=\frac{3}{2 \varrho}  \sum_{j \in \Delta}
\left( u_{j,\iota}^0 u_{j,\kappa}^0 \right)
-\frac{\delta_{\iota \kappa}}{2};
\label{eqSRQ}
\end{equation}
the above tensor turns out to be diagonal, i.e.
\begin{equation}
Q_{\iota \kappa}^0 = \delta_{\iota\kappa} q_{\kappa},~
q_{\kappa} =  P_2(N_{\kappa}). 
\end{equation}
The eigenvalue with the largest magnitude (to be denoted by $\overline{q}$)
ranges between $-\tfrac12$ and $+1$, defines the nematic second--rank order parameter, and its corresponding
eigenvector defines the nematic director $\mathbf{n}$
\cite{r23,r24,rbpz-02}.

Some specific configurations and their corresponding $\overline{q}$ quantities 
are
\begin{subequations}
\label{D-block}
\begin{align}
	D_1 = & D\left(0,\Phi \right),~\forall \Phi, & \overline{q}  & =  +1
\label{D-block-01}
\\
D_2 = & D\left(\tfrac\pi2,\tfrac\pi4 \right), & \overline{q} &  = -\tfrac12
\label{D-block-02}
\\
D_3 = & D\left(\arccos\left(\tfrac1{\sqrt{3}}\right),\tfrac\pi4
\right), & \overline{q} & =  0;
\label{D-block-03}
\end{align}
\end{subequations}
other equivalent cases can be obtained from Eqs. \eqref{D-block} by 
appropriate choices of the two angles, 
corresponding to a suitable relabeling
of lattice axes; for example, there are
six possible $D_1$-type configurations, corresponding to
${\bg u}^0(0,0,0)$ being oriented in opposite senses along
a lattice axis [{\it i.e.}
${\bg u}^0(0,0,0)=\pm {\bg e}_{\alpha},~{\alpha}=1,2,3$].

As for geometric aspects of Eq. (\ref{D-block}),
in $D_1-$type configurations, all unit vectors $\mathbf{u}^0_j$
are oriented along
a lattice axis, with appropriate signs of the corresponding components,
\textit{i.e.} a spin sitting at a lattice site and, say, its vertical neighbors
point in the same sense, its horizontal nearest neighbors point in the 
opposite way, then its horizontal next--nearest neighbors point in the same way, {\it etc}, and here full nematic order is realized.
On  the other hand, 
in $D_2-$type configurations, all unit vectors $\mathbf{u}^0_j$
lie on a lattice plane,
and their components along the corresponding  axes are
$(\pm \sqrt{2}/2,\pm \sqrt{2}/2)$, with the four combinations of signs,
producing antinematic order; finally,
in $D_3-$type configurations, the unit vectors 
$\mathbf{u}^0_j$
have components  along lattice axes given by
$(\pm \sqrt{3}/3,\pm \sqrt{3}/3,\pm \sqrt{3}/3)$, with all possible 
combinations of signs; in the latter case,
magnetic order of the unit vectors  $\mathbf{w}^0_j$
is accompanied by no nematic order;
the three named ground--state configurations can be seen 
in  FIG. 1 of Ref. \cite{pre90022506}.
Notice also that, upon integrating over the two angles, the three
quantities $q_{\kappa}$ are averaged to zero;
in other words, {\it the ground--state possesses ferromagnetic order with respect to the $\mathbf{C}^0-$ vectors, but its degeneracy destroys overall nematic order.}

According to available 
mathematical results \cite{rFroh,jsp1341059}, 
overall magnetic order (in terms of $\mathbf{C}$ vector) survives
at suitably low but finite temperatures; 
on the other hand,
different $D$ configurations might  be affected by fluctuations
to different extents, possibly to the
extreme situation
where only some of them are thermally selected (``survive'');
this behavior, studied in a few  cases after 1980, is known
as ordering by disorder, see, e.g. Refs.
\cite{od00,od01,od02,od03,od04,od05,od06}.

Actually, our additional simulations, presented in Section 
\ref{results}, showed evidence of nematic order by disorder:
it was observed that simulations started at low temperature from 
different configurations $D\left( \Theta,~\Phi \right)$
quickly resulted in configurations
remaining close to the above $D_1$ type, 
i.e. the ${\bg C}-$vector remained aligned with a lattice axis;
this caused the onset of second--rank nematic order,
as shown by  sizable values of the corresponding order
parameters  $\overline{P}_2$ and $\overline{P}_4$; in turn, the  
nematic director remained aligned with the above ${\bg C}-$vector
(see following sections); thus simulation results will suggest that, in the 
low--temperature regime, 
the above six $D_1$--type configurations correspond to pure Gibbs 
states.

\section{Computational aspects}\label{comptaspect}
Calculations were carried out using periodic boundary conditions, and  
on samples consisting of $N = L^3$ particles, with
$L=10,12,16,20,24$.
Simulations, based on standard Metropolis updating algorithm, were
carried out in cascade, in order of increasing temperature $T$,
starting at $T=0.01$; equilibration
runs took between 25000 and 50000 cycles (where one cycle
corresponds to $N$ attempted Monte Carlo steps, and production runs 
took between 500000 and 2000000; 
the Ewald--Kornfeld method with tin--foil (conducting) boundary
conditions was used for calculating configuration potential
energy \cite{mp03200931,rAT,rFS}. 

Individual attempts were carried out by first randomly selecting a lattice site,
followed by a selection of a lattice axis,
and finally carrying out a random rotation of the selected particle around it;
this algorithm was introduced by Barker and Watts some time ago \cite{cpl00300144,rAT}.

The Ewald--Kornfeld formulae  for  the potential energy 
of a given configuration of dipoles contain  
both a  pairwise summation over the direct lattice (usually truncated by the
nearest--image convention), and a sum over reciprocal lattice 
vectors (whose number is independent of $N$), essentially based on  single--particle terms \cite{mp03200931,rAT,rFS};
evaluating the energy variation resulting from the attempted random
rotation of a selected particle requires considering interactions with
the remaining $(N-1)$ particles, as well as a sum over the named
reciprocal lattice vectors: additional tests had shown  that the
computational effort requested by our program for attempting some
large number of cycles (the same for different values of $N$) scaled
with $N$ like a linear combination $(a_1 N + a_2 N^2)$.

As for calculated  thermodynamic and structural properties, as well as
finite--size scaling (FSS) analysis, we closely followed Ref.
\cite{pre90022506};  the procedure for characterizing nematic
orientational order is also reported in the Appendix.

Calculated quantities include 
the potential energy $U$ in units $\epsilon$ per particle,
and configurational specific heat $C_V/k_B$; as in Ref.
\cite{pre90022506}, we use $\mathbf{C}$ to denote
the staggered magnetization vector of a configuration, $\mathbf{m}$ for
the corresponding unit vector, 
$M$ for mean staggered magnetization, and $\chi$ for  the corresponding
susceptibility \cite{rchi1,rchi2};

We also calculated    
the fourth--order Binder cumulant $U_L$ of the staggered magnetization
\cite{pre90022506}, as well as second-- and fourth--rank
nematic order parameters $\overline{P}_2$ and $\overline{P}_4$
\cite{r23,r24,rbpz-02}, by analyzing one configuration every cycle
(see also Appendix for their definitions);
the fourth-order cumulant,
also known as the Binder cumulant \cite{rchicum02} is defined by
\begin{equation}
\label{cumuleq}
U_L=1-\frac{\langle \left(\mathbf{C} \cdot \mathbf{C} \right)^2\rangle}{3\langle\mathbf{C} \cdot \mathbf{C} \rangle ^2} .
\end{equation}
 
Correlation
between staggered magnetization and even--rank orientational
order \cite{pre90022506} was also investigated;
for a given configuration, let $\mathbf{n}$ denote the nematic
director \cite{r23,r24,rbpz-02},
and let $\mathbf{m}$ be the unit vector defined by $\mathbf{C}$;
thus we calculated the quantity
\begin{equation}
\phi = \langle | \mathbf{m}\cdot \mathbf{n}| \rangle,
\label{eqphi}
\end{equation}
where $\phi$ ranges between $\tfrac12$ for random mutual
orientation of the two unit vectors, and 1 when they are
strictly parallel or antiparallel \cite{pre90022506}.
 
\section{Results}\label{results}
\begin{figure}[!ht]
{\centering\includegraphics[scale=0.42]{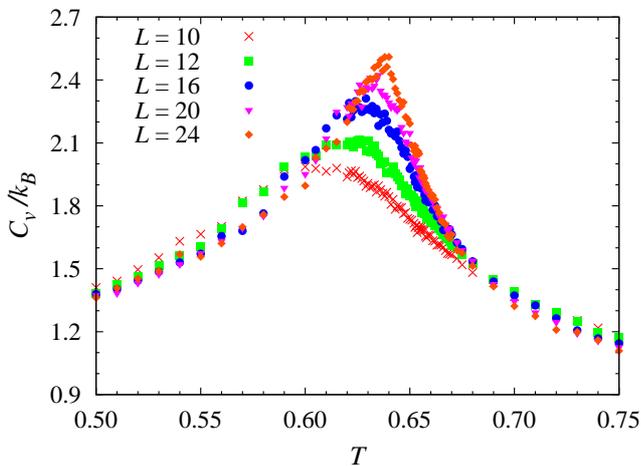}}
\caption{(Color online)
Simulation results for the configurational specific heat,
obtained with different sample sizes $L$;
the statistical errors (not shown) range between 1
and 5\%.
Meaning of symbols: red crosses: $L=10$; green squares: $L=12$;
blue circles: $L=16$; magenta triangles: $L=20$; red diamonds: $L=24$.
}
\label{figcv}
\end{figure}

Simulations estimates of the potential energy per spin (not shown here) 
were found to vary in a gradual and continuous fashion against
temperature and seemed to be largely 
unaffected by sample size to within
statistical errors ranging up to $0.5\%$.
In addition, they exhibited a smooth change of
slope at about $T\approx 0.65$. This change is reflected on the
behavior of the specific heat, whose fluctuation results showed a
recognizably  size dependent maximum around the same temperature; the
height of the maximum increases and the
``full width at half maximum'' decreases as the system size increases 
(FIG. \ref{figcv}); this behavior seems to  develop
into a singularity in the infinite--sample limit.
\begin{figure}[!ht]
{\centering\includegraphics[scale=0.42]{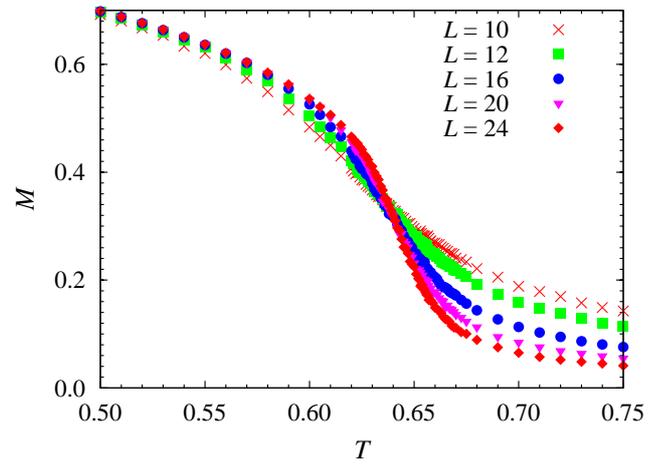}}
\caption{(Color online)
Simulation estimates for the mean staggered magnetization $M$,
obtained with different sample sizes; here and in the following figures,
the errors fall within symbol size; same meaning of symbols 
as in FIG. \ref{figcv}.
}
\label{figmag}
\end{figure}

As in our previous paper \cite{pre90022506},  and as anticipated 
in Section \ref{ground}, analysis of simulation results 
showed that, in the ordered region, the staggered magnetization vector 
$\mathbf{C}$ remains  aligned to a lattice main axis:
for example, at $T \le 0.05$, the component of $\mathbf{m}$
largest in magnitude was found to be $\ge 0.95$.
As mentioned  in the Introduction,  a spin wave treatment 
predicts orientational order at finite temperature, and the prediction
was later mathematically justified in \ci{jsp1341059}: the present
simulation results are consistent with a spin wave picture
of low--temperature excitations.

\begin{figure}[!ht]
{\centering\includegraphics[scale=0.42]{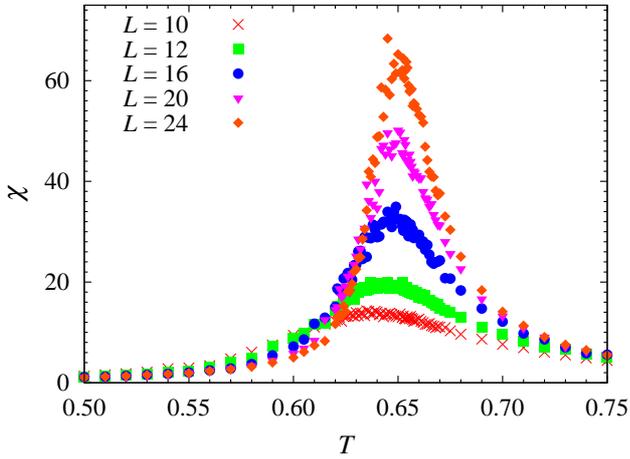}}
\caption{(Color online)
Simulation estimates for the susceptibility $\chi$ associated with
the staggered magnetization $M$, obtained with different sample sizes;
same meaning of symbols as in FIG. \ref{figcv}.}
\label{figsusc}
\end{figure}

Results for the mean staggered magnetization $M$, plotted in FIG.
\ref{figmag}, were found to decrease with temperature at fixed sample
size. For temperatures below 0.5 the data for different sample sizes 
practically coincide, while for larger temperatures the magnetization 
decreases significantly as the system size increases. The fluctuations 
of $M$ versus temperature are investigated trough the susceptibility 
$\chi$, shown in FIG. \ref{figsusc}. We observed a 
pronounced growth of 
this quantity with the system size at about $T=0.65$. This is 
manifested by a significant increase in the maximum height, as well as
a shrinking of the ``full width at half maximum'', suggesting that the
susceptibility will show a singularity as the system size goes to
infinity. This behavior is an evidence of the onset of a second order
phase transition.

To extract the critical behavior of our model a detailed FSS analysis
was applied first to the simulation data obtained for the staggered
magnetization $M$ (FIG. \ref{figmag}). This aimed at collapsing all
simulation measurements into a single curve describing
the behavior of the corresponding scaling function according to the scaling law
\begin{equation}
M=L^{-\beta/\nu}\Theta_M\left(tL^{1/\nu}\right);
\label{fssmag}
\end{equation}
here 
$t=1-\tfrac{T}{T_c} \ll 1$ denotes the distance from the bulk critical
temperature $T_c$, $\beta>0$ the critical exponent related to $M$ in
the bulk limit i.e. $\lim_{L\to\infty}M\sim t^\beta$,
and  $\nu$ is the critical exponent
for the correlation length $\xi$, \textit{i.e.}
$\xi\sim t^{-\nu}$; 
the function $\Theta_M(x)$ is a universal
function depending on the gross features of the system, but not on its
microscopic details.

\begin{figure}[!ht]
{\centering\includegraphics[scale=0.42]{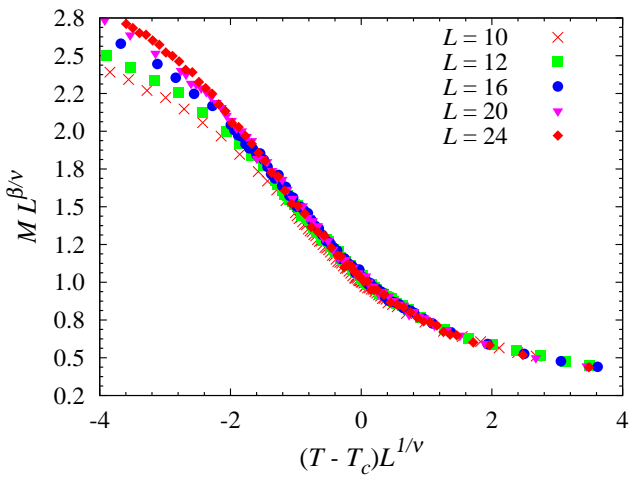}}
\caption{(Color online)
Scaling behavior of the staggered magnetization $M$;
same meaning of symbols as in FIG. \ref{figcv}.}
\label{collapse_mag}
\end{figure}

To get the best estimates for the critical exponents, several attempts
have been made on different sets of sample sizes following closely the
procedure explained in Ref. \cite{pre90022506},
which is based on the minimization approach of Ref.
\cite{melchert_2009}. The quality of
the fit was controlled by a parameter $S$ that was found to range
between the values $1$ and $2$ for all quantities considered below.
The behavior of the resulting scaling function for the staggered
magnetization is reported in FIG.
\ref{collapse_mag} with the critical temperature
$T_c=0.655\pm0.001$ and critical exponents $\beta=0.38\pm0.03$ and
$\nu=0.69\pm0.03$.

A similar analysis was performed on the simulation data for the 
susceptibility $\chi$ leading to $T_c=0.655\pm0.001$ and critical
exponents $\gamma=1.31\pm0.05$ and $\nu=0.74\pm0.01$. Here we
anticipate that, due the large fluctuations of the susceptiblity in the
vicinity of the critical temperature (see e.g. FIG. \ref{figsusc}),
 this result may be  incorrect. The fitting
procedure was attempted on the specific heat as well resulting in
$T_c=0.652\pm0.003$, $\alpha=0.13\pm0.02$ and $\nu=0.69\pm0.03$.
These results indicate that accounting for the dipolar full-range
interaction affects both nonuniversal quantities, such as the
critical temperature, and  universal features, \textit{i.e.} critical
exponents of various thermodynamic quantities. It is worth mentioning
that a similar behavior is
found in spin systems with algebraically decaying long-range
interactions of ferromagnetic type
(see e.g. Ref. \cite{chamati_critical_2003} and  references
therein also covering the bulk case).

\begin{figure}[!ht]
{\centering\includegraphics[scale=0.42]{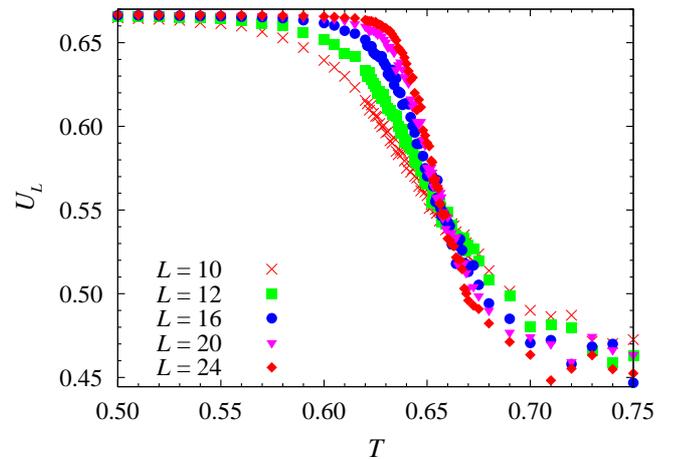}}
\caption{(Color online)
Simulation results for the fourth-order Binder cumulant
of the staggered magnetization \cite{pre90022506}
obtained with different sample 
sizes; same meaning of symbols 
as in FIG. \ref{figcv}.}
\label{cumul}
\end{figure}

Simulation estimates for the fourth-order Binder cumulant $U_L$
are shown on FIG.
\ref{cumul}. The plots for the different curves are found to decrease
against the temperature and to intersect at about $T=0.65$. A FSS
of this quantity yields the critical temperature to a very good
approximation, since a data collapse leads a scaling function that is
independent on the sample size. This is found to be
$T_c=0.656\pm0.002$ and the critical exponent $\nu=0.69\pm0.08$.
At the critical temperature we obtain the critical amplitude
$U_L^*\approx0.54$.

\begin{figure}[!ht]
{\centering\includegraphics[scale=0.42]{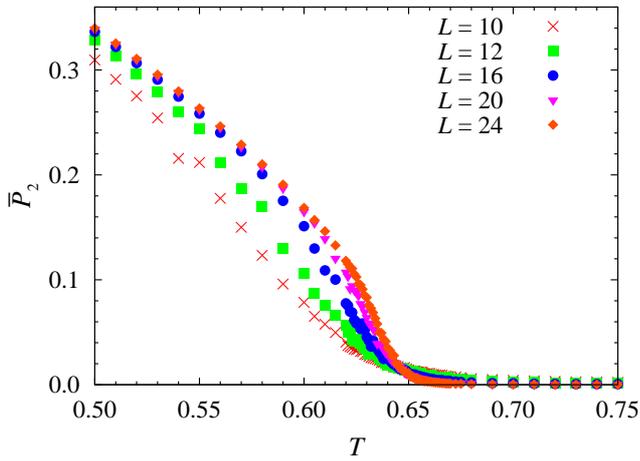}}
\caption{(Color online)
Simulation results
for the nematic second--rank order parameter $\overline{P}_2$,
obtained with different sample sizes:
same meaning of symbols as in FIG. \ref{figcv}.}
\label{figp2}
\end{figure}

\begin{figure}[!ht]
{\centering\includegraphics[scale=0.42]{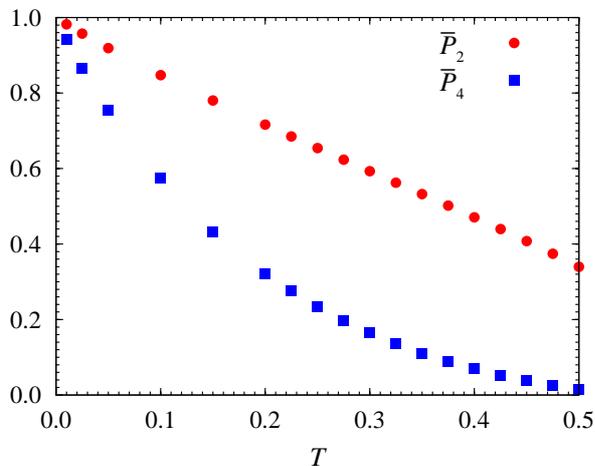}}
\caption{(Color online) 
Comparison between simulation results
for nematic second-- and fourth--rank order parameters, in the low--temperature region, 
and obtained with the largest investigated sample size $L=24$;
meaning of symbols: red circles: $\overline{P}_2$;
blue squares: $\overline{P}_4$}
\label{figQQ}
\end{figure}

At all investigated temperatures, 
simulation results for the {\em nematic} order parameters
$\overline{P}_2$ 
and $\overline{P}_4$ (FIGs. \ref{figp2} and  \ref{figQQ})
exhibited a gradual and monotonic  decrease with temperature, 
vanishing  above $T_c$, and appeared  to be mildly affected by sample sizes;
results for  $\overline{P}_4$ became negligible
in the transition region, $T \gtrsim 0.55$ (not shown); 
in the low--temperature region, simulation results for both
observables tended  to saturate to 1 as  $T \rightarrow  0^{+}$.

According to FSS approach the 
nematic order parameter is expected to scale like
\begin{equation}
\overline{P}_2=L^{-2\beta/\nu}\Xi\left(tL^{1/\nu}\right).
\end{equation}
Applying the above mentioned minimization procedure we get 
$T_c=0.655 \pm 0.002$, 
$\beta=0.37\pm0.02$ and $\nu=0.69\pm0.03$ in a 
very good
agreement with the above finding for the staggered magnetization.


\begin{figure}[ht!]
{\centering\includegraphics[scale=0.42]{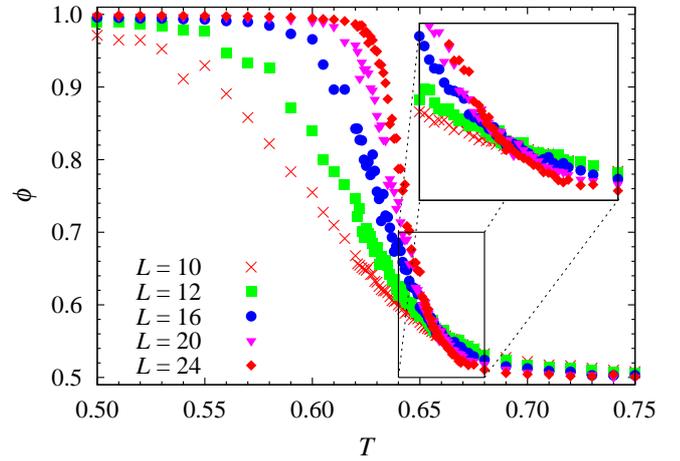}}
\caption{(Color online)
Simulation results
for the quantity $\phi$, as defined in the text (Eq. \ref{eqphi}), 
obtained with different sample sizes; same meaning of symbols as in 
FIG. \ref{figcv}.}
\label{figrho}
\end{figure}

Simulation data for $\phi$ [Eq. (\ref{eqphi})] are plotted
in FIG. \ref{figrho}; for all investigated sample sizes they
appear to decrease with increasing temperature;
moreover, the results
exhibit a recognizable  increase of $\phi$
with increasing sample size for $T \lesssim T_1 = 0.64$, 
and its recognizable decrease
with increasing sample size for $T \gtrsim T_2 = 0.68$,
so that the seemingly continuous change  across the transition region
becomes steeper and steeper as sample  size increases.
In the crossover temperature range between $T_1$ and $T_2$ 
the sample--size dependence of results becomes
rather weak, and the various curves come close to coincidence at
$T\approx 0.66 \pm 0.01$, with $~\phi \approx 0.54$; notice
that this temperature value is in reasonable agreement with $T_c$ as 
independently estimated  via the above FSS
treatment.

Let us recall that, by  construction, the quantity $\phi$
should be size
independent at the critical temperature and thus all curves should coincide
there, as it is the case for the Binder cumulant. A FSS analysis
was carried out, and found to support
this conjecture, giving results consistent with those for
$U_L$.

\begin{figure}[ht!]
{\centering\includegraphics[scale=0.42]{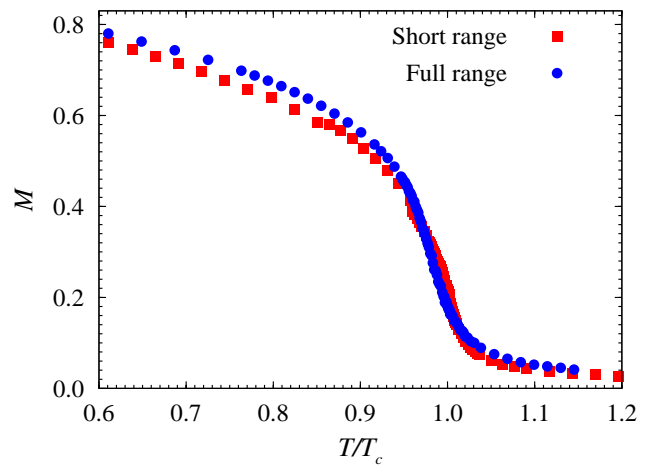}}
\caption{(Color online)
Plots of $M$ versus $T/T_c$ for  the present model and its nearest--neighbor
counterpart \cite{pre90022506}; meaning of symbols: blue circles: present 
full--range model; red squares: nearest--neighbor counterpart.
Both here and in the following FIG. \ref{figcompp2}, the plotted simulation results
had been obtained for $L=24$.} 
\label{figcompfm}
\end{figure}

\begin{figure}[ht!]
{\centering\includegraphics[scale=0.42]{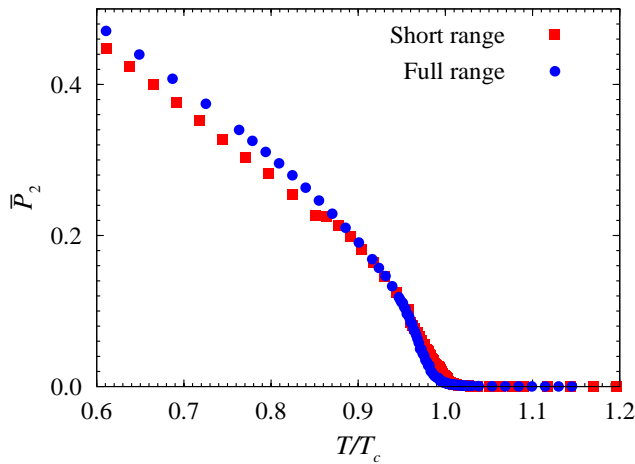}}
\caption{(Color online)
Plots of  $\overline{P}_2$
versus $T/T_c$ for  the present model and its nearest--neighbor
counterpart \cite{pre90022506}; same meaning of symbols: 
as in FIG. \ref{figcompfm}.} 
\label{figcompp2}
\end{figure}

To summarize, we propose for the critical
temperature the value $T_c=655\pm0.005$, versus the corresponding 
value $1.877 \pm 0.001$ in
Ref. \cite{pre90022506}, thus the
ratio $\rho = T_c/\left|W_{GS}\right|$ drops to roughly one half
of its short--range counterpart ($\approx 0.24$ versus $\approx 0.470$),
and this suggests that the long--range tail of the interaction
reduces the stability range of the ordered phase in comparison
with the nearest--neighbor case.
Comparison with the short--range counterpart was also realized 
in FIGs. \ref{figcompfm} and \ref{figcompp2}, where
simulation results for M and $\overline{P}_2$,  obtained with  the largest  
sample--size used in both studies ($L=24$),
are plotted versus $T/T_c$;
the Figures show a pronounced similarity,
as well as a mild but  recognizable strengthening of orientational
order in the low--temperature region.

On the other hand the critical behavior was found to be governed by
the critical exponents $\nu = 0.69\pm0.08$, $\beta =
0.38\pm0.03$ and $\alpha = 0.13 \pm 0.02$. 
Except for the above result 
$\gamma = 1.31 \pm 0.05$, these values  
are in agreement with previous  
Renormalization Group (RG) calculations for isotropic
dipolar criticality (Table I in  Ref.
\cite{bruce_critical_1974}), as well as the
experimental measurements of
Ref. \cite{ravi_kumar_magnetic_2015} on Cr$_{70}$Fe$_{30}$,
obtained on films of appropriate thickness. Since the value
of $\gamma$ is highly affected by large fluctuations in  the
critical region,
in order to  get a meaningful result we employed the hyperscaling
relations to obtain $\gamma\approx1.37$. Thus the model investigated
here  is consistent with the isotropic dipolar universality class;
comparison between transitional properties for the present
model and for its nearest--neighbor counterpart is summarized
in the following Table.

\begin{table}[ht!]
\caption{Comparison between transitional properties for the present model 
and its nearest--neighbor counterpart \cite{pre90022506}.\label{T01}}
\begin{ruledtabular}
\begin{tabular}{lcc}
~~~ & Present model & Ref.~\cite{pre90022506}
\\
\hline
$W_{GS}$  & $-2.676$ & $-4$   
\\                                                       
$T_{c}$   & $0.655 \pm 0.005$ &  $1.877\pm 0.001$ 
\\
$\rho=\tfrac{T_c}{\left|W_{GS}\right|}$ &  $\approx 0.24$ &  $\approx 0.470$
\\
$\alpha$ & $0.13 \pm  0.02$ &  $0.13 \pm 0.02$
 \\
$\beta$  & $0.38 \pm 0.03$ & $0.358 \pm 0.006$
\\
$\gamma$ & $\approx 1.37$ & $1.4 \pm 0.1$
\\
$\nu$ & $0.69 \pm 0.08$ & $0.713 \pm 0.001$
\\
\end{tabular}
\end{ruledtabular}
\end{table}

As for comparison with other treatments, let us first mention that a
Weiss--type Molecular Field approach predicts a transition temperature
$T_{c,MF}=\tfrac23\left|W_{GS}\right|$, {\it i.e.} $\tfrac83 \approx 2.667$ for 
the nearest--neighbor counterpart, and $1.784$ in the present case
\cite{johnston_magnetic_2016,nuovocimd7}, hence
the ratio $T_c/T_{c,MF}=\tfrac32 \rho$ has dropped by the same numerical factor 
(nearly 2) as above.

Both nearest--neighbor and full--ranged
cases of the model investigated here had been studied some sixty
years ago by the spherical model (SM) approach
\cite{rsm01,rsm02,rsm03,rsm04,rsm05}; as
for the nearest--neighbor case, the estimated transition temperature
can be obtained from Eq. (32) in \cite{rsm03} by 
evaluating a multiple integral numerically, 
{\it i.e.} $T_{c,SM} = 1.693$.
In the full--ranged case, as far as we could check,
Ref. \cite{rsm02} did not report any explicit numerical 
estimate of the transition temperature in their Eq. (6.1);
on the other hand, some results are available in Ref. \cite{rsm05},
via their Eq. (1.3) (with their $\alpha$ set to $0$) and following treatment
[see also their FIGs. (1) and (2)];
these results read $F(\lambda_M) \approx 0.73$, and hence 
$T_{c,SM} \approx 0.457$.

The critical exponents reported in the above papers
were $\beta = \tfrac12$
and $\alpha = 0$ for both cases: in both cases the configurational
specific heat $C_V/k_B$ was found to remain constant at $\tfrac32$ for
$T \le T_{c,SM}$, and to change
continuously but with a discontinuous slope at $T = T_{c,SM}$.

An interaction model defined by an extension of \eqref{e01}
had later been studied in \cite{pssb09800307} by RG;
the interaction potential was defined by 
\begin{eqnarray}
W_{ij}  &=&  \epsilon [
-(1+2 \sigma)(3+2 \sigma) (\mathbf{u}_i \cdot \hat{\mathbf{r}}_{ij})
(\mathbf{u}_j \cdot \hat{\mathbf{r}}_{ij})
\nonumber
\\ & &+ (1+2\sigma)(\mathbf{u}_i \cdot 
\mathbf{u}_j )]/r^{3+\sigma},~
\label{e01gen}
\end{eqnarray}
where $\sigma \ge 0$ is a real parameter: 
all critical exponents with the exception of $\beta$ were found
to depend on $\sigma$, and 
the limiting case $\sigma=0$
corresponded to the model studied here, for which Eqs. (47) in the
named paper \cite{pssb09800307} yield
$\alpha=1$, $\beta=\tfrac12$, $\nu=1$, $\eta=2$, $\gamma=0$.

\section{Conclusions}\label{conclusions}
We have studied here
the transitional behavior resulting from the full--ranged counterpart of 
the lattice--spin
model in Refs. \cite{pre90022506,prb4912287}, by means of
simulation as well as a detailed analysis of results;
FSS basically suggests a universality class
with critical exponents $\nu = 0.69\pm0.08$, $\beta =
0.38\pm0.03$ and $\alpha = 0.13 \pm 0.02$,
and a critical temperature $T_c=0.655 \pm0.005$,
{\it i.e.} consistent with an isotropic dipolar critical point
\cite{ravi_kumar_magnetic_2015,bruce_critical_1974} 
and different
from the nearest--neighbor ferromagnetic Heisenberg one; 
analysis of second--rank properties has shown the existence of secondary
nematic order, destroyed by ground state degeneracy but
restored the low--temperature phase,
through a mechanism of order by disorder 
(see Ref. \cite{pre90022506} and others quoted therein).

The ratio $\rho = T_c/\left| W_{GS} \right|$  drops to roughly one half
of its short--range counterpart ($\approx 0.24$ versus $\approx 0.470$)
and this suggests that the long--range tail of the interaction
reduces the stability range of the ordered phase
in comparison with the nearest--neighbor case.

\section*{Acknowledgements}
The present extensive calculations were carried out, on,
among other machines,
workstations, belonging to the Sezione di Pavia of
Istituto Nazionale di Fisica Nucleare (INFN); allocations of computer
time by the Computer Centre of Pavia University and CILEA
(Consorzio Interuniversitario Lombardo
per l'Elaborazione Automatica, Segrate - Milan),
as well as by CINECA
(Centro Interuniversitario Nord-Est di Calcolo Automatico,
Casalecchio di Reno - Bologna),
and CASPUR (Consorzio interuniversitario per le 
Applicazioni di Supercalcolo per Universit\`a e Ricerca, Rome)
are gratefully acknowledged.

\appendix*
\section{Nematic second-- and fourth--rank order parameters} \la{secondandfourth}
Both second-- and fourth--rank
nematic order parameters \ci{r23,r24,rbpz-02} were calculated by analyzing one
configuration every cycle;  in other words, 
for a generic examined configuration, the ${\sg Q}$ tensor 
is defined by the appropriate generalization of Eq. \eqref{eqSRQ},
now involving all the spins in the sample, {\it i.e.}
\begin{equation}
Q_{\iota \kappa}= \tfrac12(3 F_{\iota \kappa} - \delta_{\iota \kappa}),
\label{e1801}
\end{equation}
with
\begin{equation}
F_{\iota \kappa}= \langle u_{\iota} u_{\kappa} \rangle_{loc}=
\frac{1}{N} \sum_{j=1}^N \left( u_{j,\iota} u_{j,\kappa}\right);
\end{equation}
here $\langle \ldots \rangle_ {loc}$  denotes average
over the current configuration.
The fourth-rank order parameter was determined via  the analogous 
quantity \cite{RL13}
\begin{eqnarray}
B_{\iota \kappa \lambda \mu}&=& \frac{1}{8}
[35 G_{\iota \kappa \lambda \mu}
- 5 ( \delta_{\iota \kappa} F_{\lambda \mu}
+ \delta_{\iota \lambda} F_{\kappa \mu} +
\delta_{\iota \mu} F_{\kappa \lambda}
\nonumber\\ & &
+\delta_{\kappa \lambda} F_{\iota \mu}
+\delta_{\kappa \lambda} F_{\iota \mu} + 
\delta_{\lambda \mu} F_{\iota \kappa} )
\nonumber\\ & &
+(\delta_{\iota \kappa} \delta_{\lambda \mu} +
\delta_{\iota \lambda}\delta_{\kappa \mu} + 
\delta_{\iota \mu} \delta_{\kappa \lambda} ) ],
\label{e19}
\end{eqnarray}
where
\begin{equation}
G_{\iota \kappa \lambda \mu}= 
\langle u_{\iota} u_{\kappa} u_{\lambda} u_{\mu} \rangle_{loc}
= \frac{1}{N} \sum_{j=1}^N 
u_{j,\iota} u_{j,\kappa} u_{j,\lambda} u_{j,\mu}.
\end{equation}
The calculated tensor ${\sg Q}$ was diagonalized;
let $\omega_k$ denote its three eigenvalues, and let
$\mathbf{v}_k$ denote the corresponding eigenvectors;
the  eigenvalue with the 
largest magnitude (usually a positive number, thus the maximum eigenvalue),
can be identified, and its 
average over the simulation chain
defines the nematic second--rank  order parameter
$\overline{P}_2$;   the corresponding
eigenvector defines the local (fluctuating   or  ``instantaneous'')
configuration director $\mathbf{n}$ \cite{r23,r24,rbpz-02}, evolving
along the simulation.  Moreover, a 
suitable reordering of eigenvalues (and hence of the corresponding 
eigenvectors) is needed  for evaluating $\overline{P}_4$;
let the eigenvalues $\omega_k$ be reordered (permuted
according to some rule), to yield
the values $\omega_k^{\prime}$; 
the procedure used here as well as
in other previous papers (e.g. Refs. \cite{rcont1,rcont2}) involves a permutation
such that
\begin{subequations}
\la{eqreorder}
\begin{equation}
|\omega_3^{\prime}| \ge |\omega_1^{\prime}|,
~|\omega_3^{\prime}| \ge |\omega_2^{\prime}|;
\label{eqper1}
\end{equation}
actually there exist two such possible permutations, an odd and an even one;
we consistently chose permutations of the same
 parity (say  even ones, see also below)
for all examined configurations; 
recall that eigenvalue reordering also induces the corresponding 
permutation of the associated eigenvectors.
Notice also that, in most cases, 
$\omega_3^{\prime} >0$, so that the condition in Eq. (\ref{eqper1})
reduces to
\begin{equation}
\omega_3^{\prime} \ge \omega_1^{\prime},
~\omega_3^{\prime} \ge \omega_2^{\prime};
\la{eqper2}
\end{equation}
\end{subequations}
this latter procedure was considered  in earlier treatments of the
method. As already mentioned,
the second--rank order parameter
$\overline{P}_2$ is defined by the average of $\omega_3^{\prime}$
over the simulation chain; on the other hand,
the quantity $(\omega_2^{\prime}-\omega_1^{\prime})$,
and hence its average over the chain, measure  possible phase
biaxiality, 
found here to be zero within statistical errors, as it
should. The procedure outlined here was previously used
elsewhere \cite{rcont1,rcont2,r091,r093,r096,r097},
in cases where some amount of
biaxial order might exist; the consistent choice
of permutations of the same parity was found to avoid
both artificially enforcing  a spurious phase biaxiality 
(as would result by imposing an
additional condition such as
$|\omega_1^{\prime}| \le |\omega_2^{\prime}|$ ), and
artificially reducing or even quenching it
(as would result by ordering $\omega_1^{\prime}$ and
$\omega_2^{\prime}$ at random).

The fourth-rank order parameter was evaluated from the ${\sg B}$
tensor in the 
following way \cite{RL13}: for each analyzed configuration, 
the suitably reordered eigenvectors of ${\sg Q}$ 
define the director frame, and build the
column vectors
of an orthogonal matrix ${\sg R}$, in turn employed
for transforming 
${\sg B}$ to the director frame;   the diagonal element
$B_{3333}^{\prime}$ of the transformed tensor 
was  averaged over the production run, and identified 
with $\overline{P}_4$. 

%
\end{document}